\documentclass[aps,twocolumn,superscriptaddress,showpacs,amsmath]{revtex4}

\usepackage{epsf,latexsym}
\usepackage{amssymb,amsmath}
\usepackage{times}

\newcommand{\ket}[1]{| #1 \rangle}

\newcommand\R{{\mathrm {I\!R}}}

\newcommand\diag{{\mbox{diag\,}}}

\newcommand{\ignore}[1]{}

\newcommand{\ra}{{\rightarrow}}
\newcommand{\up}{{\uparrow}}
\newcommand{\down}{{\downarrow}}

\newcommand{\be}{\begin{equation}}
\newcommand{\ee}{\end{equation}}
\newcommand{\ba}{\begin{eqnarray}}
\newcommand{\ea}{\end{eqnarray}}

\def\CC{{\rm\kern.24em \vrule width.04em height1.46ex depth-.07ex
    \kern-.30em C}}
\def\P{{\rm I\kern-.25em P}}

\def\RR{{\rm
         \vrule width.04em height1.58ex depth-.0ex
         \kern-.04em R}}

\def\bbbc{{\mathchoice {\setbox0=\hbox{$\displaystyle\rm C$}\hbox{\hbox
to0pt{\kern0.4\wd0\vrule height0.9\ht0\hss}\box0}}
{\setbox0=\hbox{$\textstyle\rm C$}\hbox{\hbox
to0pt{\kern0.4\wd0\vrule height0.9\ht0\hss}\box0}}
{\setbox0=\hbox{$\scriptstyle\rm C$}\hbox{\hbox
to0pt{\kern0.4\wd0\vrule height0.9\ht0\hss}\box0}}
{\setbox0=\hbox{$\scriptscriptstyle\rm C$}\hbox{\hbox
to0pt{\kern0.4\wd0\vrule height0.9\ht0\hss}\box0}}}}

\def\bbbz{{\mathchoice {\hbox{$\sf\textstyle Z\kern-0.4em Z$}}
{\hbox{$\sf\textstyle Z\kern-0.4em Z$}}
{\hbox{$\sf\scriptstyle Z\kern-0.3em Z$}}
{\hbox{$\sf\scriptscriptstyle Z\kern-0.2em Z$}}}}

\newcommand{\putfig}[2]{$$\leavevmode\hbox{\epsfxsize=#2 cm
   \epsffile{#1.eps}}$$}

\begin{document}

\title{Generalized Toffoli gates using qudit catalysis}

\author{Radu Ionicioiu}
\affiliation{Hewlett-Packard Laboratories, Long Down Avenue, Stoke Gifford, Bristol BS34 8QZ, UK}

\author{Timothy P.~Spiller}
\affiliation{Hewlett-Packard Laboratories, Long Down Avenue, Stoke Gifford, Bristol BS34 8QZ, UK}

\author{William J.~Munro}

\affiliation{Hewlett-Packard Laboratories, Long Down Avenue, Stoke Gifford, Bristol BS34 8QZ, UK}
\affiliation{National Institute of Informatics, 2-1-2 Hitotsubashi, Chiyoda-ku, Tokyo 101-8430, Japan}

\begin{abstract}
We present quantum networks for a $n$-qubit controlled gate $C^{n-1}(U)$ which use a higher dimensional (qudit) ancilla as a catalyser. In its simplest form the network has only $n$ two-particle gates (qubit-qudit)-- this is the minimum number of two-body interactions needed to couple all $n+1$ subsystems ($n$ qubits plus one ancilla). This class of controlled gates includes the generalised Toffoli gate $C^{n-1}(X)$ on $n$ qubits, which plays an important role in several quantum algorithms and error correction. A particular example implementing this model is given by the dispersive limit of a generalised Jaynes-Cummings Hamiltonian of an effective spin-$s$ interacting with a cavity mode.

\end{abstract}


\maketitle

\section{Introduction}

During the past decade quantum information processing (QIP) has changed our paradigm of computation \cite{qip}. The main insight was that information is physical, and hence computation, as a physical process, is limited by the laws of physics \cite{deutsch}. Quantum mechanics provides a different computational model by employing features not present in a classical world, most notably superposition and entanglement. The quantum revolution opened new possibilities by devising novel algorithms \cite{shor, grover} and tasks which either have no classical analog \cite{teleport, densecode, crypto}, or provide (in some cases) an exponential speed-up over classically known algorithms \cite{shor}.

A critical issue that needs to be addressed is how to construct large scale quantum circuits. As classical computing is designed around irreversible gates, one cannot directly translate this expertise into the quantum world. However, reversible logic has a long history \cite{ft, toffoli} and a central role in this field is played by the Toffoli gate. This gate, a controlled-controlled-NOT acting on three bits, flips the target bit if both controls are 1. There are design tools that allow to construct large oracles with reversible gates.

The Toffoli gate is also of interest in the quantum world. Together with the Hadamard it forms a universal set of quantum gates \cite{shi}. Moreover, the Toffoli gate is a central building block in phase estimation \cite{nc}, Shor's algorithm \cite{shor}, error correction \cite{cory} and fault tolerant quantum circuits \cite{denis}. These properties naturally bring up the question of how to construct it efficiently. Since the Toffoli involves 3-body interactions, it does not appear naturally in physical systems, with Hamiltonians usually containing 2-body interactions. One way of constructing the Toffoli is to decompose it into single and two qubit gates, the simplest decomposition in terms of CNOT gates require 6 such gates plus 10 single qubit gates; equivalently, one can use five two-qubit gates, but in this case these are general controlled-$U$ instead of CNOT \cite{barenco}. One can generalize Toffoli gates to $n$ qubits, $n-1$ controls and a target, where now the target qubit is flipped if all the controls are 1. For generalized Toffoli gates the resources increase rapidly, requiring ${\cal O}(n^2)$ two-qubit gates \cite{barenco}. Experimentally, 3-qubit Toffoli gates have been implemented in NMR systems \cite{cory} and ion traps \cite{toffoli_ion}.

Recently a new approach has provided a fresh insight of how to construct efficiently Toffoli gates \cite{ralph, lanyon} using fewer resources than previous designs \cite{nc, barenco}. The key element in this construction \cite{ralph, lanyon} is the use of a larger Hilbert space by transforming the target qubit into a qudit. Intuitively, expanding the two dimensional space of a qubit to a $n$ dimensional qudit space improves the ``manoeuvrability'' in the extended Hilbert space and results in a simplified quantum network with fewer gates. However, this requires that during the gate operation the state of the system goes out of the computational space -- effectively one of the computational qubits becomes a qudit. Therefore, at the end of the gate, this qudit has to re-enter the qubit space in order to continue the computation and hence needs to ``forget'' the non-computational degrees of freedom. A consequence is the potential leakage out of the qubit space: imperfect gates will result in leakage, especially if such a procedure is reiterated during a complex algorithm.

In this article we build on this insight \cite{ralph, lanyon} but adopt a different approach resulting in a more efficient design. We completely decouple the qudit space from the computational space by using an ancilla. This has several important consequences. First, there is no need to expand one of the computational qubits to a qudit and then back to a qubit. Second, this separation of the ancilla from the computational degrees of freedom allows us to measure and discard the ancilla at the end of the gate. Consequently we recover the computational Hilbert space in a pristine form -- the ancilla starts and ends in a factorized state with respect to the qubits, hence there is no leakage. Here measurement plays also the role of a simple error correction mechanism. And finally, we have the freedom to implement the qudit in a different physical system than the computational qubits. This means we can optimize other properties of the ancilla compared to the qubits (e.g., interaction strength, higher dimensional Hilbert space etc). For example, the computational qubits can be photons (low decoherence) and the qudit ancilla can be an atom in a cavity (higher decoherence but stronger interaction); in this case the ancilla has to be coherent only for the duration of the Toffoli gate, not for the duration of the whole algorithm. The ability to measure the ancilla proves to be crucial and enables us to reduce the number of entangling gates: our second design needs only $n$ two-particle gates, compared to $2n-3$ in Ref.~\cite{ralph,lanyon}. Note that $n$ is the minimum number of two-body interactions required to couple $n+1$ subsystems ($n$ qubits plus one ancilla): less than this, at least one of them is left uncoupled with the rest of the system.

We will discuss two related designs. The first does not require measurement, but needs $2n-1$ two-body gates. The second design uses only $n$ two-particle gates plus measurement of the ancilla and a feed-forward correction (single qubit phase shifts on the control qubits). It is important to stress that both networks are deterministic and the result of the measurement (in the second case) is {\em not} used for post-selection. Here measurement plays the same role as in teleportation: it provides the information needed to correct the final state -- the gate works with unit probability in principle. Finally, we give an example of a particular model which can be used to implement our scheme. This corresponds to the dispersive limit of a generalized Jaynes-Cummings Hamiltonian of an effective spin-$s$ interacting with a cavity mode.

\section{Quantum network for Toffoli gate}

\subsection{Background and notations} 

Before discussing the main result of the article, let us briefly review a few background ideas and notations. A qudit is a $n$ level quantum system; let $\{ \ket{0}, \ldots, \ket{n-1} \}$ be its $n$ dimensional (computational) basis. The action of the generalized Pauli operators $X_n$ and $Z_n$ in this basis is:
\begin{eqnarray}
X_n \ket{k}&=& \ket{k+1} \cr
Z_n \ket{k}&=& \omega^k \ket{k}
\label{xz}
\end{eqnarray}
with $\omega= e^{2\pi i/n}$; the addition in $\ket{k+1}$ is modulo $n$. Thus $Z_n= \diag(1, \omega, \omega^2, \ldots, \omega^{n-1})$ in this basis. For $n=2$, we denote the usual Pauli operators acting on a qubit by $X$ and $Z$. The Fourier transform over the qudit space is defined as \cite{nc}:
\be
F \ket{k}= n^{-1/2} \sum_{j=0}^{n-1} \omega ^{kj} \ket{j}
\label{ft}
\ee
The Fourier transform maps the computational $Z_n$-basis to the $X_n$-basis (up to a relabeling); we also have $F X_n F^{-1}= Z_n$. For a qubit ($n=2$) $F$ is the Hadamard gate $H$.

A quantum gate $C^{n-1}(U)$ is defined to have $n-1$ control qubits and performs a conditional $U$ gate on the $n$-th qubit (the target) if and only if all the controls are 1, i.e., $i_1\cdot \ldots \cdot i_{n-1}=1$, with $i_k=0,1$. The action is given by:
\be
C^{n-1}(U) \ket{i_1 \ldots i_n}= \ket{i_1 \ldots i_{n-1}} U^{i_1\cdot \ldots \cdot i_{n-1}}\ket{i_n}
\ee
The case $U=X$ corresponds to the generalized Toffoli gate acting on $n$ qubits, namely $C^{n-1}(X)$; its action is to flip the target qubit if all the $n-1$ control qubits are in the $\ket{1}$ state.

\subsection{A first gate design}

Now we are ready to discuss the quantum network for a generalized controlled gate $C^{n-1}(U)$ using a $n$-dimensional qudit as an ancilla.

\begin{figure}
\putfig{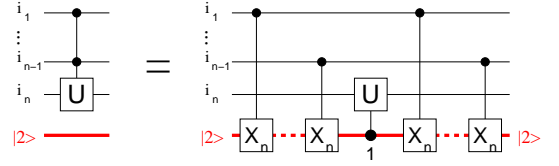}{7}
\caption{An $n$-qubit controlled $U$ gate $C^{n-1}(U)$; if $U=X$ this corresponds to a generalized Toffoli gate. Qubits $i_1, \ldots, i_n$ are denoted by thin, black lines; the ancilla (red, thick line) is a qudit with a $n$ dimensional Hilbert space. The controlled-$U$ gate in the middle between a qubit and the qudit ancilla is performed only if the ancilla is in the $\ket{1}$ state, i.e., $C(U) \ket{i_n}\ket{j}= [U^{\delta_{j1}}\ket{i_n}] \ket{j}$ (symbolized by 1 below the gate).}
\label{t1}
\end{figure}

Consider the quantum network in Fig.~\ref{t1}. The controlled-$U$ gate in the middle between the qudit ancilla (control) and $i_n$ (target) is performed only if the qudit is in the $\ket{1}$ state, i.e., $\ket{i_n}\ket{j} \ra [U^{\delta_{j1}}\ket{i_n}] \ket{j}$. With this observation, the gate operation becomes straightforward. For basis states, the ancilla acts like a counter, adding all the values of the control qubits. Since the initial state of the ancilla qudit is $\ket{2}$, the middle $C(U)$ gate is performed when $2+i_1+ \ldots +i_{n-1}= 1 \mod n$, that is all control qubits are in the state $\ket{1}$. The last $n-1$ gates $C(X_n)$ after the $C(U)$ gate disentangle the ancilla from the computational qubits and return it to its initial state.

The choice of the initial state $\ket{2}$ for the ancilla is determined by the fact that the middle $C(U)$ gate (between the ancilla and the $n$th qubit) acts only if the ancilla is in the $\ket{1}$ state. If, experimentally, we want to have the $C(U)$ gate controlled by the $\ket{k}$ value, one has simply to prepare the ancilla in the $\ket{k+1}$ state.

\begin{figure}
\putfig{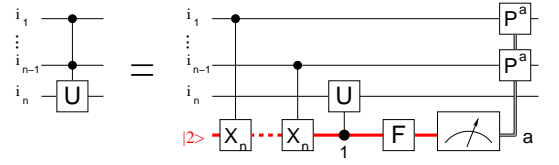}{7}
\caption{A measurement based $C^{n-1}(U)$ gate. The $n-1$ gates $C(X_n)$ used to disentangle the ancilla from the qubits are replaced by a measurement of the ancilla in the Fourier basis (gate $F$ on the qudit) and feed-forward. The gates $P^a= \diag(1, \omega^a)$ are conditionally applied depending the value $a$ of the measured ancilla; the same gate $P^a$ is applied to all control qubits, resulting in a homogeneous design.}
\label{t2}
\end{figure}

\subsection{Disentangling by measurement: a simpler design}

In the previous scheme each control qubit has to interact twice with the qudit, first to entangle, then to disentangle it from the ancilla (see Fig.~\ref{t1}). If the qubits are photons and the ancilla is an atom in a cavity, as in the photonic module \cite{ph_module, ph_module2}, then all the control photons have to pass twice through the cavity. From an experimental point of view this will require to store the photons in a memory or buffer and then redirect them back to the cavity, a situation far from optimal. Is there a way to simplify the network such that each qubit interacts only once with the ancilla?

In Fig.~\ref{t2} we present an alternative design. By measuring the ancilla in the Fourier basis (i.e., by first applying a Fourier transform $F$ followed by a measurement in the computational basis) and using feed-forward, we can effectively disentangle the ancilla from the rest of the qubits.

Assume we have an arbitrary input state $\ket{\psi_0}= \sum_i \alpha_i \ket{i_1 \ldots i_n} \ket{2}$, where the sum is over all basis states $i=(i_1, \ldots, i_n)$ of $n$ qubits; the ancilla is factorized and starts in the state $\ket{2}$. After applying the $n-1$ controlled-$X_n$ gates $C(X_n)$ on the ancilla, the state becomes $\ket{\psi_1}= \sum_i \alpha_i \ket{i_1 \ldots i_n} \ket{i_1+ \ldots + i_{n-1}+ 2}$ and the ancilla is now entangled with all the control qubits $i_1, \ldots, i_{n-1}$. By applying the $C(U)$ gate between the ancilla and the target qubit $i_n$, the state of the system changes to
\[
\ket{\psi_2}= \sum_i \alpha_i \ket{i_1 \ldots i_{n-1}} U^{i_1\cdot \ldots \cdot i_{n-1}}\ket{i_n} \ket{i_1+ \ldots + i_{n-1}+ 2}
\]
After the Fourier transform $F$ on the ancilla, eq.(\ref{ft}), the previous state becomes (up to normalization):
\[
\sum_i \alpha_i \ket{i_1 \ldots i_{n-1}} U^{i_1\cdot \ldots \cdot i_{n-1}}\ket{i_n} \sum_{k=0}^{n-1} \omega^{k(i_1+ \ldots + i_{n-1}+ 2)} \ket{k}
\]
Suppose we now measure the ancilla and we obtain the value $a$, with $0\le a \le n-1$; the system is then projected to the state
\[
\sum_i \alpha_i \ket{i_1 \ldots i_{n-1}} U^{i_1\cdot \ldots \cdot i_{n-1}}\ket{i_n} \omega^{a(i_1+ \ldots + i_{n-1}+ 2)} \ket{a}
\]
Hence by applying a corrective phase shift $P^a$ to all $n-1$ control qubits, with $P= \diag(1, \omega)$, we obtain
\[
\ket{\psi_f}= \omega^{2a} \sum_i \alpha_i \ket{i_1 \ldots i_{n-1}} U^{i_1\cdot \ldots \cdot i_{n-1}}\ket{i_n} \ket{a}
\]
which is indeed the desired state, modulo an (irrelevant) overall phase $\omega^{2a}$.

It is important to stress that although the ancilla is measured, there is no post-selection, hence the gate is deterministic and works with unit probability. Measurement here plays the same role as in teleportation -- it provides the information we need to correct the state in the end.

\subsection{A generalized Jaynes-Cummings model}

\begin{figure}
\putfig{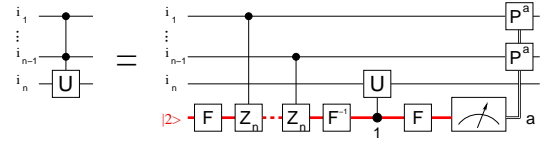}{7}
\caption{A quantum network equivalent to Fig.~\ref{t2}, using the identity $X_n=F^{-1} Z_n F$.}
\label{t3}
\end{figure}

So far the discussion was rather general without considering a particular physical model. We now examine a possible implementation of the abstract scheme presented above. The first question we ask is: {\em What interaction Hamiltonian gives us the controlled unitary transformation $C(X_n)$?} Using the identity $X_n=F^{-1} Z_n F$, the problem reduces to find a Hamiltonian whose evolution enacts a controlled-$Z_n$ gate $C(Z_n)$, as in Fig.~\ref{t3}.

Assume we have a qudit (e.g., an atom in a cavity or a nuclear spin in a NMR setup) interacting with a photonic qubit according to
\be
H_I= \chi \hbar\, a^\dag a\, S_z
\label{hi}
\ee
where $a^\dag (a)$ are photonic creation (annihilation) operators and $S_z= \diag(s, s-1, \ldots, -s+1, -s)$ is the effective $z$-spin operator associated to the qudit, with $n= 2s+1$ the dimension of its Hilbert space. The unitary operator induced by this (time independent) Hamiltonian acting for a time $t$ on the system is
\be
U(t)= e^{-iH_I t/\hbar}= \exp(-i \chi t a^\dag a S_z)
\ee
Let $\ket{k}\ket{m}$ be a basis state of the system, with $\ket{k}, k=0,1$ a Fock state of the field (photon) and $\ket{m}$, $0\le m \le n-1$, a basis for the qudit. The action on this basis is
\be
U(t) \ket{k} \ket{m}= \ket{k} [\tilde U^k(t) \ket{m}]
\ee
with $\tilde U(t)= \exp(-i \chi t S_z)$ a single qudit unitary. Taking $\chi t=2\pi /n$ and recalling that $\omega=e^{2\pi i/n}$, we obtain
\be
\tilde U(2\pi/\chi n)= -\omega^{1/2} \diag(1, \omega, \ldots, \omega^{n-1})= -\omega^{1/2} Z_n
\ee
Therefore the Hamiltonian (\ref{hi}) implements, up to a phase, the unitary we need.

It is well-known that in the case of a qubit ($n=2$) the Hamiltonian $H_I$ can be obtained as the dispersive limit of the Jaynes-Cummings Hamiltonian $H_{JC}= \hbar g(a^\dag \sigma_- + a \sigma_+)$. This Jaynes-Cummings Hamiltonian \cite{JC,GerKni} describes the interaction of an {\em effective} spin-1/2 (e.g., a 2-level atom in a cavity) with a quantized field.

It is easy to show that the dispersive limit holds also for the general case general spin-$s$. Assume we have a spin-$s$ (or equivalent) in a cavity interacting with a quantized field according to:
\be
H= \hbar \omega a^\dag a + \hbar \Omega S_z + 2 \hbar g (a^\dag S_- + a S_+)
\label{eq:genJC}
\ee
where $\omega$ ($\Omega$) is the resonant frequency of the cavity field (spin) and $S_\pm= S_x \pm i S_y$ are the ladder operators for a spin $s$, with $n=2s+1$. These obey the usual commutation relations $[S_+, S_-]= 2 S_z$ and $[S_z, S_\pm]= \pm S_\pm$. The dispersive limit $g/\Delta \ll 1$, with $\Delta= \Omega- \omega$ the detuning, is obtained by performing the unitary transformation \cite{cavityQED} $V= \exp[2g(aS_+- a^\dag S_-)/\Delta]$ and expanding to second order in $g$
\begin{eqnarray}
H'&=& VHV^\dag= \hbar \omega a^\dag a + \hbar \Omega S_z + \frac{4\hbar g^2}{\Delta} (S_+ S_- + 2 a^\dag a S_z) \cr
&+&{\cal O}(g^3/\Delta^2) 
\end{eqnarray}
This points towards two possible routes for physical implementation. One possibility is the dispersive limit ($g/\Delta \ll 1$) of an actual spin $s$ in a cavity, interacting with the field through the Hamiltonian of eq.~(\ref{eq:genJC}) and with capability for $S_z$ measurement. An alternative is to use $n-1$ dispersive qubits, each at the same frequency $\Omega$, not coupled to each other but all coupled to the field with the same strength \cite{TC}. With
\be
S_z = \frac{1}{2} \sum_{j=1}^{n-1} \sigma_{z}^{(j)} \;\;\;\;\;\;\;\;
S_{\pm} = \frac{1}{2} \sum_{j=1}^{n-1} \sigma_{\pm}^{(j)}
\label{eq:n-1spin}
\ee
the required commutation rules hold for the effective spin $s$ (where $n=2s+1$) and the Hamiltonian of eq.~(\ref{eq:genJC}) emerges. Clearly here the full $(n-1)$-qubit Hilbert space is bigger than that required. However, since the evolution preserves the symmetry, if the qubits are initialised in the symmetric (under qubit interchange) subspace they remain in it and act as an effective spin $s$. For the two-qubit case $n=3$, the symmetric states, written in terms of the individual $\sigma_z$ eigenstates would be $\{\ket{\up \up},\frac{1}{\sqrt{2}}\left(\ket{\up \down}+\ket{\down \up}\right),\ket{\down \down} \}$. Given the symmetric subspace, measurement projecting into the $S_z$ basis can be effected by measuring the individual qubits in their $\sigma_z$ basis.

The other core ingredient of Fig.~\ref{t3} is the Fourier transform (FT) over the qudit. It can be implemented efficiently with at most $n^2-1$ pulses -- any unitary $U\in SU(n)$ can be written as \cite{zhou} $U= \prod_{k=1}^{n^2-1} e^{i\beta_k T_k}$, with $T_k$ the $SU(n)$ generators and $\beta_k \in \R$. Proposals to implement the FT include multilevel atoms \cite{QFT_qudit} and harmonic oscillators \cite{bartlett}.

\section{Conclusions}

A challenge for the present day quantum engineers is to minimize the quantum resources required to perform a given quantum task. In this article we discussed two quantum networks implementing a family of controlled gates on $n$ qubits $C^{n-1}(U)$. This includes the generalized Toffoli gate $C^{n-1}(X)$, an important ingredient of quantum error correction algorithms.

Recently several authors proposed a new approach for an efficient implementation of generalized Toffoli gates \cite{ralph, lanyon}. The insight was to use a higher dimensional Hilbert space by allowing one of the qubits (the target) to temporarily become a qudit during the gate operation. The end result is a reduction in the number of two-body quantum gates required at the expense of going out and in of the computational space during gate execution.

Let us now review the main features of our construction. Our design strategy was to decouple the higher dimensional space (the qudit) from the computational qubits by choosing the qudit to be an ancilla. This has several consequences: first, it prevents leakage from the computational space, as the ancilla can be discarded at the end of the gate. Second, the qubits and the ancilla can be implemented in different physical systems (e.g., photons vs.~an atom in a cavity). And third, as the ancilla does not have any computational degrees of freedom, it can be measured without destroying the coherence of the qubits. This is the case of the gate design in Fig.~\ref{t2}, where measuring the ancilla in the Fourier basis disentangles it from the computational qubits and simplifies the design: each qubit interacts minimally with the ancilla -- only once. Another advantage of the present scheme is that all control qubits interact homogeneously, i.e., in the same way, with the qudit; moreover, there is no need of extra single qudit gates on the ancilla between the $C(X_n)$ gates (in contrast, in Ref.~\cite{ralph, lanyon} extra single-qudit gates are required). This last property is especially useful for quantum computation schemes in which the ancilla is symmetrically coupled to all control qubits. One particular example is a NMR approach where a central qubit is coupled identically (due to symmetry) to a number of satellite qubits, as in Ref.~\cite{NMR_NOON}. In this case we can enact the $n-1$ $C(X_n)$ gates between the control qubits and the ancilla with a {\em single} NMR pulse, provided the target qubit is distinct from the rest, hence has a different coupling.

The controlled $C(Z_n)$ unitary we use in our scheme can be implemented with a Hamiltonian which generalizes the dispersive limit of Jaynes-Cummings interaction of a qudit coupled to a photonic field. One physical implementation of this system could be an actual matter qudit (a spin $s$, with $n=2s+1$) coupled to a cavity field. Another possibility is to realise an effective qudit from $n-1$ qubits each coupled to the field. The latter approach could be feasible with superconducting qubits each positioned at an antinode of a microwave cavity field \cite{TCexpt}. For this example the required measurement projecting into the $S_z$ basis might be effected through dispersive measurement of the individual qubits \cite{readout1,readout2} at their operating points, or possibly through gate voltage bias change followed by charge measurement.

\acknowledgments

We acknowledge financial support from EU (QAP and HIP projects) and Japan (MEXT).


\end{document}